\documentclass[12pt]{article}%
\usepackage{citesort}
\usepackage{amsmath}
\usepackage{graphicx}%
\usepackage{amsfonts}%
\usepackage{amssymb}
\begin{document}

\begin{center}
\bigskip

\bigskip

\textbf{{\large Central inclusive dijets production by double pomeron
exchange.}}

\textbf{{\large Comparison with the CDF results}}
\end{center}

\vskip                                                           3mm

\begin{center}
Adam Bzdak
\end{center}

\vskip                                                           3mm

\begin{center}
M. Smoluchowski Institute of Physics, Jagiellonian University \\
Reymonta 4, 30-059 Krak\'{o}w, Poland

E-mail: \texttt{bzdak@th.if.uj.edu.pl}
\end{center}

\vskip                                                  .5cm

\begin{quotation}
Using the Landshoff-Nachtmann model of the po{\-}meron, rapidity and
transverse momentum distributions of gluon jets produced by double po{\-}meron
exchange in \textit{pp }(\textit{p\={p}}) collisions are calculated. The
comparison with the CDF Run I results on the central inclusive dijets
production cross sections is performed. We find the model to give correct
order of magnitude for the measured cross sections.

\bigskip

\end{quotation}

\section{Introduction}

The subject of Higgs boson production by double pomeron exchange (DPE) has
drawn noticeable interest in recent years
\cite{Schafer,Bial-Land,Cudell,Levin,Khoze-Higgs-Jets,Khoze-Higgs,ET4,Cox,Peschan-G,Peschan,Ingelman}%
.

The lack of solid QCD framework for diffraction makes the determination
difficult. Despite the distinct progress in the recent years the serious
uncertainties are still present.

One way to reduce these uncertainties is to study other double pomeron
exchange processes and compare them with existing data. A particularly
illuminating process is the DPE production of two jets (dijets). Such a
process was originally discussed at the Born level in \cite{Barera}. Later the
dijets production was studied in \cite{Khoze-Higgs-Jets,Koze-Jets} and in
\cite{Cox,Peschan-G,Peschan,Ingelman,Forshaw-jets}.

Recently, using the Landshoff-Nachtmann model of the pomeron, the
cross-section for gluon jets production was calculated \cite{Bzdak}. The
obtained results together with those for quark-antiquark jets calculated some
time ego \cite{Bial-Szer,Bial-Janik} give the full cross-section for dijet
production in double pomeron exchange reactions\footnote{It is generally
accepted that DPE dijets mainly consist of gluon jets. For that reason we will
use \textit{gluon jets} and \textit{dijets} interchangeably.}.

In the present paper we calculate the differential cross-section for gluon
jets production hoping that this will allow a more precise comparison of the
model with experiment.

The comparison with the CDF Run I results on the central inclusive dijets
production cross sections \cite{CDF-1} is also performed. \begin{figure}[h]
\begin{center}
\includegraphics[width=6cm,height=4cm]{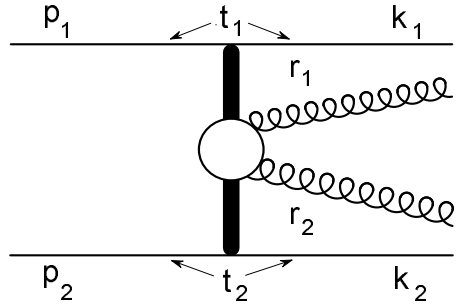}
\end{center}
\caption{Production of two gluons by double pomeron exchange.}%
\label{ogolny}%
\end{figure}

\section{Matrix element and notation}

In the Landshoff-Nachtmann model \cite{Land-Nacht} the pomeron is approximated
by an exchange of two non-perturbative gluons coupled to one of the quarks of
the colliding hadrons.

\begin{figure}[h]
\begin{center}
\includegraphics[width=3.80cm,height=2cm]{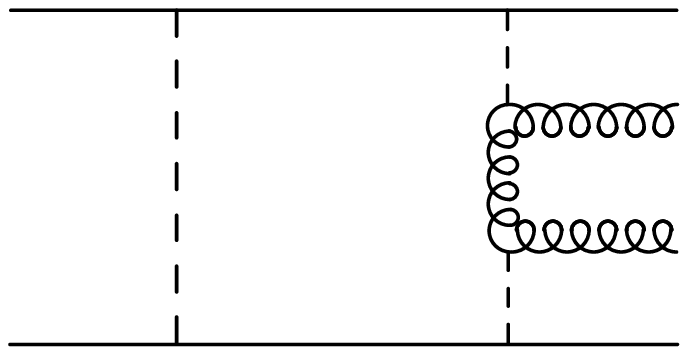}  \hspace{0.3cm}
\includegraphics[width=3.80cm,height=2cm]{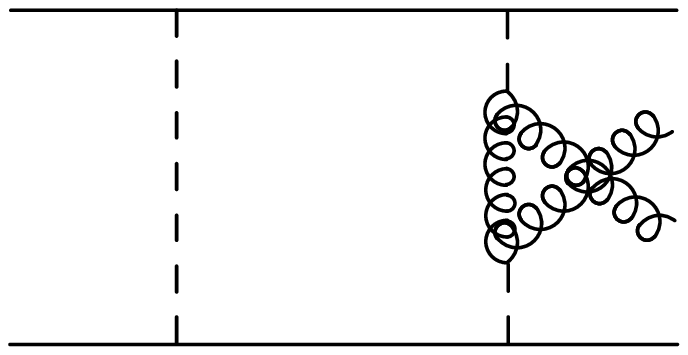}  \hspace{0.3cm}
\includegraphics[width=3.80cm,height=2cm]{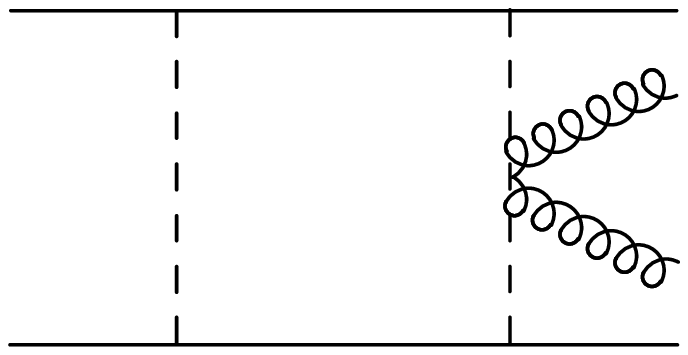}
\end{center}
\caption{Three diagrams contributing to the amplitude of the process of gluon
pair production by double pomeron exchange. The dashed lines represent the
exchange of the non-perturbative gluons.}%
\label{ok}%
\end{figure}

The matrix element for gluon jets by double pomeron exchange in such model is
given by the sum of the three diagrams shown in Fig. \ref{ok} (plus analogous
emissions from the second gluon)

\noindent where the inner quark lines are put on shell. In this model, it was
shown that the square of the matrix element (averaged and summed over spins
and polarizations) for the diffractive production of two gluons is of the form
\cite{Bzdak}:%
\begin{equation}
\overline{\left|  M_{pp}\right|  ^{2}}=81\overline{\left|  M_{qq}\right|
^{2}}\left[  F\left(  t_{1}\right)  F\left(  t_{2}\right)  \right]  ^{2},
\label{pp}%
\end{equation}
where $\overline{\left|  M_{qq}\right|  ^{2}}$ is the dijet production
amplitude squared for colliding quarks\footnote{Note the Regge-like dependence
implied by Eq. ($\ref{qq}$). There are some controversies if such assumption
is justified. The resent results \cite{Peschan-G,Peschan} on the inclusive DPE
dijets production, however, are quite encouraging.}$^{\text{,}}$\footnote{It
is worth to stress that formula (\ref{qq}) is only valid in the limit of
$\delta_{1,2}<<1$ and for small momentum transfer between initial and final
quarks.}:%
\begin{equation}
\overline{\left|  M_{qq}\right|  ^{2}}=C\frac{s^{2}}{\left(  u_{_{1}}\right)
^{2}\left(  u_{_{2}}\right)  ^{2}}\delta_{1}^{2-2\alpha(t_{1})}\delta
_{2}^{2-2\alpha(t_{2})}\exp\left(  2\beta\left(  t_{1}+t_{2}\right)  \right)
R^{2}. \label{qq}%
\end{equation}
Transverse momenta of the produced gluons are denoted by $u_{_{1}}$ and
$u_{_{2}}$. The constants $C$ and $R$ are defined later. $\alpha\left(
t\right)  =1+\epsilon+\alpha^{\prime}t$ is the pomeron Regge trajectory with
$\epsilon\approx0.08,$ $\alpha^{\prime}=0.25$ GeV$^{-2}$ ($t_{1},$ $t_{2}$ are
marked in Fig. \ref{ogolny}). $F\left(  t\right)  $ = $\exp($ $\lambda t)$ is
the nucleon form-factor with $\lambda=$ $2$ GeV$^{-2}$. $\delta_{1},$
$\delta_{2}$ are defined as $\delta_{1,2}\equiv1-k_{1,2}/p_{1,2}$ ( $k_{1},$
$k_{2},$ $p_{1},$ $p_{2},$ are marked in Fig. \ref{ogolny}). The factor
$\exp\left(  2\beta\left(  t_{1}+t_{2}\right)  \right)  $ with $\beta$ $=$ $1$
GeV$^{-2}$ \cite{Factor} takes into account the effect of the momentum
transfer dependence of the non-perturbative gluon propagator given by ($p^{2}$
is the Lorentz square of the momentum carried by the non-perturbative gluon):
\begin{equation}
D\left(  p^{2}\right)  =D_{0}\exp\left(  p^{2}/\mu^{2}\right)  .
\label{propag}%
\end{equation}

The constants $C$ and $R$ are defined as:%
\begin{align}
C  &  =\frac{1}{\left(  27\pi\right)  ^{2}}\left(  D_{0}G^{2}\mu\right)
^{6}\mu^{2}\left(  \frac{g^{2}/4\pi}{G^{2}/4\pi}\right)  ^{2},\label{consP}\\
R  &  =9\int d\vec{Q}_{\intercal}^{2}\text{\thinspace}\vec{Q}_{\intercal}%
^{2}\exp\left(  -3\vec{Q}_{\intercal}^{2}\right)  =1. \label{constR}%
\end{align}

Here $G$ and $g$ are the non-perturbative and perturbative quark gluon
couplings respectively. $\mu$ is the range of the non-perturbative gluon
propagator (\ref{propag}) and $D_{0}$ its magnitude at vanishing momentum
transfer. From data on the elastic scattering of hadrons one infers
$D_{0}G^{2}\mu\approx30$ GeV$^{-1}$ and $\mu\approx1$ GeV.

The constant $R$ shows the structure of the loop integral. $\vec{Q}%
_{\intercal}$ is the transverse momentum carried by each of the three
non-perturbative gluons.

Fig. \ref{ogolny} seems to describe exclusive dijets production. In fact, as
was clearly stated in the original paper on the Higgs production
\cite{Bial-Land}, our calculation really is a central inclusive one $i.e.$ the
radiation is present in the central region of the rapidity.

\section{ Differential cross-section}

Having the matrix element (\ref{qq}) we can write the formula for the
differential cross-section:%
\begin{equation}
d\sigma=\left(  2s\right)  ^{-1}\left(  2\pi\right)  ^{-8}\overline
{|M_{pp}|^{2}}dPH, \label{diffc-s}%
\end{equation}
where $dPH$ is a differential phase-space factor:%
\begin{align}
dPH  &  =d^{4}k_{1}\delta\left(  k_{1}^{2}\right)  d^{4}k_{2}\delta\left(
k_{2}^{2}\right)  d^{4}r_{1}\delta\left(  r_{1}^{2}\right)  d^{4}r_{2}%
\delta\left(  r_{2}^{2}\right) \nonumber\\
&  \times\Theta\left(  k_{1}^{0}\right)  \Theta\left(  k_{2}^{0}\right)
\Theta\left(  r_{1}^{0}\right)  \Theta\left(  r_{2}^{0}\right)  \delta
^{(4)}\left(  p_{1}+p_{2}-k_{1}-k_{2}-r_{1}-r_{2}\right)  . \label{faza}%
\end{align}

Expression (\ref{diffc-s}) is to be integrated over all variables except
rapidities and transverse momenta of the produced gluons. Following closely
the method used in \cite{Bial-Janik} we obtain the following result for the
differential cross-section:
\begin{equation}
\frac{d\sigma}{d^{2}u_{+}d^{2}udy_{1}dy_{2}}=C_{E}\left(  u_{_{1}}\right)
^{-2}\left(  u_{_{2}}\right)  ^{-2}\left(  \delta_{1}\delta_{2}\right)
^{-2\epsilon}\frac{\pi}{L_{1}+L_{2}}\exp\left(  -\frac{L_{1}L_{2}}{L_{1}%
+L_{2}}(u_{+})^{2}\right)  . \label{end1}%
\end{equation}
In the above expression $y_{1,2}$ are the rapidities of the produced gluons,
$u_{+}=u_{1}+u_{2}$, $u=(u_{1}-u_{2})/2$, $C_{E}=C\frac{81}{16\left(
2\pi\right)  ^{8}}$ and $L_{1,2}=2\left(  \beta+\lambda-\alpha^{\prime}%
\ln\delta_{1,2}\right)  $. $\delta_{1},$ $\delta_{2}$ are expressed by
rapidities and transverse momenta as follows:%
\[
\delta_{1}\sqrt{s}=|u_{1}|\exp\left(  y_{1}\right)  +|u_{2}|\exp\left(
y_{2}\right)  ,
\]%
\begin{equation}
\delta_{2}\sqrt{s}=|u_{1}|\exp\left(  -y_{1}\right)  +|u_{2}|\exp\left(
-y_{2}\right)  . \label{KL1}%
\end{equation}

The differential cross-section (\ref{end1}) gives a Gaussian cut-off on the
total transverse momentum of the produced pair. Putting $u_{+}=0$ $i.e$
$u_{1}=-u_{2}=u$ everywhere except in the exponent and performing the
integration over $(u_{+})^{2}$ we obtain:
\begin{equation}
\frac{d\sigma}{d(u^{2})dy_{1}dy_{2}}=C_{E}\left(  u\right)  ^{-4}\left(
\delta_{1}\delta_{2}\right)  ^{-2\epsilon}\frac{\pi^{3}}{L_{1}L_{2}}.
\label{end2}%
\end{equation}
Taking into account (\ref{KL1}) and the definition of $L_{1,2}$ we finally
obtain:%
\begin{align}
\frac{d\sigma}{d(E_{\intercal}^{2})d(\Delta y)dy}  &  =C_{E}\left(
E_{\intercal}\right)  ^{-4}\left(  \frac{4E_{\intercal}^{2}}{s}\cosh
^{2}(\frac{\Delta y}{2})\right)  ^{-2\epsilon}\nonumber\\
&  \times\frac{\pi^{3}/4\alpha^{\prime}{}^{2}}{\left(  \frac{\lambda+\beta
}{\alpha^{\prime}}-\frac{1}{2}\ln\left(  \frac{4E_{\intercal}^{2}}{s}\cosh
^{2}(\frac{\Delta y}{2})\right)  \right)  ^{2}-y^{2}}. \label{end3}%
\end{align}
Here $\Delta y=y_{1}-y_{2},$ $y=(y_{1}+y_{2})/2.$ $E_{\intercal}=$
$|u_{1}|=|u_{2}|$ is the transverse energy of one of the produced gluons.

This completes the calculations of the differential cross-section.

Some comments are in order.

(i) Since $y^{2}<<\left(  \frac{\lambda+\beta}{\alpha^{\prime}}\right)
^{2}=144$ and $\frac{4E_{\intercal}^{2}}{s}\cosh^{2}(\frac{\Delta y}%
{2})=\delta_{1}\delta_{2}$ $<<1$ the differential cross section (\ref{end3})
depends very weekly on the sum of the dijet rapidities. It is worth to mention
that the kinematical limit of the double pomeron exchange region depends on
both the sum $y$ and the difference $\Delta y$ of the rapidities, as seen from
(\ref{KL1}).

(ii) The main uncertainty in the expression (\ref{end3}) is the value of
$G^{2}/4\pi.$ Following \cite{Bial-Land,Bial-Szer} we take it to be
\cite{G/4pi} about $1.$In fact it should be considered only as an order of
magnitude estimate.

\section{Comparison with the CDF Run I results}

The CDF collaboration has presented \cite{CDF-1} results on the central
inclusive DPE dijet production cross sections.

At Run I ($\sqrt{s}=1.8$ TeV) the cross section for the central inclusive
dijets of $E_{\intercal}>7$ GeV [$E_{\intercal}>10$ GeV] is measured to be
$43.6$ $\pm$ $4.4$(stat) $\pm$ $21.6$(syst) nb [$3.4$ $\pm$ $1$(stat) $\pm$
$2$(syst) nb]. The kinematics is following: $0.01<\delta_{1}\equiv\delta
_{p}<0.03$, $0.035<\delta_{2}\equiv\delta_{\bar{p}}<0.095$, jets are confined
within $-4.2<y<2.4$ and the gap requirement $2.4<y_{gap}<5.9$ on the proton side.

It should be noted that in the above experiment the protons were not detected
and the DPE events were enhanced by a rapidity gap requirement on the proton
side. Thus, in principle, there is no specified cuts on the outgoing
proton\footnote{In principle the result (\ref{end3}) should by multiplied by a
factor $\left(  1-\exp\left(  -L_{1}\frac{s\left(  1-\delta_{1}\right)  ^{2}%
}{\exp\left(  2y_{gap}^{\max}\right)  }\right)  \right)  $ where
$y_{gap}^{\max}$ is the maximum value of the gap. In the present case,
$y_{gap}^{\max}=5.9$, this factor is close to $1$.}.

Integrating\footnote{Note an identical final state particle phase space factor
$\frac{1}{2!}$.} (\ref{end3}) over the appropriate kinematical range we obtain
the results shown in Table \ref{CDF1-noGap}. The running coupling constant
$g^{2}/4\pi,$ appearing in (\ref{consP}), is evaluated at $2E_{\intercal
}^{\min}$ $i.e.$ $0.15$ and $0.14$ for $E_{\intercal}^{\min}=7,10$ GeV
respectively. $G^{2}/4\pi$ is taken to be $1.$ \begin{table}[h]
\begin{center}%
\begin{tabular}
[c]{|c|c|c|}\hline\hline
Transverse energy & CDF & $\sigma$\\\hline
$E_{\intercal}>7$ GeV & $43.6$ $\pm$ $26$ [nb] & $70$ [nb]\\\hline
$E_{\intercal}>$ $10$ GeV & $3.4$ $\pm$ $3$ [nb] & $30$ [nb]\\\hline
\end{tabular}
\end{center}
\caption{Comparison of the results obtained in the present paper with the CDF
results on the central inclusive DPE dijets production cross sections.
$G^{2}/4\pi$ is taken to be $1.$}%
\label{CDF1-noGap}%
\end{table}

As can be seen the original Bialas-Landshoff model \cite{Bial-Land} for DPE
diffractive production gives correct order of magnitude for the measured
central inclusive cross sections.

\section{Discussion}

The results shown in Table \ref{CDF1-noGap} do not take gap survival effect
($S_{gap}^{2}$) into account $i.e.$ the probability of the gaps not to be
populated by secondaries produced in the soft rescattering. Following
\cite{Khoze-Higgs-Jets,CDF-S2} we take it for the Tevatron energy to be about
$0.1$ (it is the best we can do). Multiplying the results shown in Table
\ref{CDF1-noGap} by $S_{gap}^{2}$ we obtain the results presented in Table
\ref{CDF1-Gap}.\begin{table}[h]
\begin{center}%
\begin{tabular}
[c]{|c|c|c|c|}\hline\hline
Transverse energy & CDF & $%
\begin{array}
[c]{c}%
\sigma\times S_{gap}^{2}\\
G^{2}/4\pi=1
\end{array}
$ & $%
\begin{array}
[c]{c}%
\sigma\times S_{gap}^{2}\\
G^{2}/4\pi=0.4
\end{array}
$\\\hline
$E_{\intercal}>7$ GeV & $43.6$ $\pm$ $26$ [nb] & $7$ [nb] & $43$ [nb]\\\hline
$E_{\intercal}>$ $10$ GeV & $3.4$ $\pm$ $3$ [nb] & $3$ [nb] & $19$
[nb]\\\hline
\end{tabular}
\end{center}
\caption{Comparison of our results with the CDF results on the central
inclusive DPE dijets production. Gap survival factor is taken into account. }%
\label{CDF1-Gap}%
\end{table}

Some comments seems to be in order.

(i) If we assume $G^{2}/4\pi=0.4$ so that we reproduce the result for
$E_{\intercal}>7$ GeV, the result for $E_{\intercal}>10$ GeV overestimates the
measured cross section about $5$ times. It means that some suppression with
$E_{\intercal}$ is needed.

(ii) The expression (\ref{end3}) (+$S_{gap}^{2}$) do not include the Sudakov
factor $i.e.$ the probability that the rapidity gaps survive QCD radiation. We
think that in our particular case, $E_{\intercal}>7,10$ GeV, the Sudakov
physics is not so essential, however, the lack of Sudakov $E_{\intercal}$
suppression can be seen. It is not obvious how the Sudakov factor should be
included in the presented model and this problem is currently under our consideration.

(iii) The factor $S_{gap}^{2}$ is not a universal number. It depends on the
initial energy and the particular final state. Theoretical predictions of the
survival factor, $S_{gap}^{2}$, can be found in Ref. \cite{S2-theory}.

(iv) We work at the partonic level. Thus, in a realistic experimental
situation our results correspond to a smaller cone where the jet axes are
confined, since the gluon produced just on the boundary of the cone usually
produces secondary particles outside the cone and such event is not counted.
This effect would decrease the calculated cross section. We checked that
taking the range of the rapidity as $-3.5<y<1.7$ (CDF $-4.2<y<2.4$) our
cross-sections decrease no more than $10\%$. To cover it fully Monte-Carlo
simulations are needed \cite{Forshaw-jets}.

Finally, let us note that estimates in the present paper, as well as in
\cite{Bial-Land,Peschan-G,Peschan,Bzdak,Bial-Szer,Bial-Janik}, are based on
the basis of the pure forward direction. It was first mentioned in
\cite{Pumplin} that such approach may lead to incorrect results.

\section{Conclusions}

In conclusion, using the Landshoff-Nachtmann model of the pomeron, we have
presented the rapidity and transverse momentum distribution of central
inclusive gluon jets produced by double pomeron exchange in \textit{pp
}(\textit{p\={p}})\textit{ }collisions.

We observed a distinct dependence on the difference of gluons rapidities
$\Delta y$ and the marginal dependence on their sum $y$. A power-law decrease
approximately as $E_{\intercal}^{-4.3}$, for relatively small $E_{\intercal}$
production when the Sudakov physics is not essential, with the increasing
transverse momentum is observed. We find the model to give reasonable
predictions, at least correct order of magnitude for the measured central
inclusive cross sections.

\bigskip

\textbf{Acknowledgements }It is a pleasure to thank Professor Andrzej Bialas
for his encouragement and to acknowledge many enlightening discussions with
Dr. Leszek Motyka. The discussions with Professor Robert Peschanski and
Professor Valery Khoze are very highly appreciated. This investigation was
supported by the Polish State Committee for Scientific Research (KBN) under
grant 2 P03B 043 24.

\end{document}